\documentclass[]{svjour2}

\smartqed

\usepackage{amssymb}
\usepackage{graphicx}

\journalname{J Stat Phys}

\begin{document}

\title{Finite-Size Effects in Non-Neutral Two-Dimensional Coulomb Fluids}

\titlerunning{Finite-Size Effects in Coulomb Fluids}

\author{Ladislav \v{S}amaj}

\institute{Institute of Physics, Slovak Academy of Sciences, 
D\'ubravsk\'a cesta 9, SK-84511 Bratislava, Slovakia \\
\email{Ladislav.Samaj@savba.sk}}

\date{Received:  / Accepted: }

\maketitle

\begin{abstract}
Thermodynamic potential of a neutral two-dimensional (2D) Cou\-lomb fluid, 
confined to a large domain with a smooth boundary, exhibits at any 
(inverse) temperature $\beta$ a logarithmic finite-size correction term
whose universal prefactor depends only on the Euler number of the domain
and the conformal anomaly number $c=-1$.
A minimal free boson conformal field theory, which is equivalent to 
the 2D symmetric two-component plasma of elementary $\pm e$ charges 
at coupling constant $\Gamma=\beta e^2$, was studied in the past.
It was shown that creating a non-neutrality by spreading out a charge 
$Q e$ at infinity modifies the anomaly number to 
$c(Q,\Gamma) = - 1 + 3\Gamma Q^2$. 
Here, we study the effect of non-neutrality on the finite-size expansion of
the free energy for another Coulomb fluid, namely the 2D one-component plasma 
(jellium) composed of identical pointlike $e$-charges in a homogeneous
background surface charge density. 
For the disk geometry of the confining domain we find that the non-neutrality 
induces the same change of the anomaly number in the finite-size expansion.
We derive this result first at the free-fermion coupling 
$\Gamma\equiv\beta e^2=2$ and then, by using a mapping of the 2D 
one-component plasma onto an anticommuting field theory formulated on a chain, 
for an arbitrary coupling constant.

\keywords{Coulomb fluid\and Finite-size correction\and Central charge\and
Conformal field theory}

\end{abstract}

\renewcommand{\theequation}{1.\arabic{equation}}
\setcounter{equation}{0}

\section{Introduction} \label{Sect.1}
According to the principle of conformal invariance 
\cite{Affleck86,Blote86,Cardy88a}, two-dimensional (2D) systems of constituents 
with {\em short-range} interactions, confined to a large domain with a smooth 
boundary, exhibit at the critical point universal finite-size properties.
In particular, for a disk domain of radius $R$,
$D = \{ {\bf r}\in \mathbb{R}^2, \vert {\bf r}\vert\le R\}$ with
the Euler number $\chi=1$, the (dimensionless) free energy has 
the large-$R$ expansion
\begin{equation} \label{finite-size}
\beta F = (\beta f) \pi R^2 + (\beta\sigma) 2\pi R - 
\frac{c}{6} \ln \left( \frac{R}{L} \right) + O(1) , 
\end{equation}
where $\beta \equiv 1/(k_{\rm B}T)$ is the inverse temperature
and $L$ is a length scale.
The bulk specific free energy $\beta f$ and the surface tension
$\beta\sigma$ are non-universal, i.e., dependent on microscopic details
of the given model and on the temperature. 
The logarithmic term is universal, depending only on the conformal 
anomaly number (central charge) $c$ of the critical theory.
Like for instance, the massless Gaussian theory with a field $\phi({\bf r})$, 
defined by the Hamiltonian 
\begin{equation} \label{GaussH}
H_{\rm G} = \frac{1}{4\pi} \int_D {\rm d}^2r\, \left( \nabla\phi \right)^2 ,
\end{equation}
has $c=1$ \cite{Cardy88b}.

In this paper, we study 2D classical (i.e., non-quantum) systems of charged 
pointlike particles interacting pairwisely by the {\em long-range} Coulomb 
potential $\phi({\bf r}) = - \ln(\vert{\bf r}\vert/L)$, where the dielectric 
constants of the medium in which particles move and of the walls are equal
to $1$ (vacuum in Gauss units), for simplicity.
The logarithmic potential is the solution of the 2D Poisson equation
\begin{equation}
\Delta \phi({\bf r}) = - 2\pi \delta({\bf r})
\end{equation}
and in three dimensions corresponds to the effective interaction of infinitely 
long parallel charged lines, perpendicular to the plane.
Two types of Coulomb systems are of special interest.
The one-component plasma (OCP), or jellium, is a system of identical 
mobile (say elementary) charges $e$ immersed in a homogeneous neutralizing 
background charge density.
The symmetric two-component plasma (TCP), or the Coulomb gas, is a system
of oppositely charged $\pm e$ mobile particles with no background.
While the OCP is usually studied in the canonical ensemble, the TCP is
treated within the grand-canonical ensemble, with the enforced neutrality
condition for each microscopic configuration.
In both cases, the thermodynamics and the particle correlation functions
are determined by the only dimensionless parameter, the coupling constant 
$\Gamma=\beta e^2$.
The 2D OCP is exactly solvable at $\Gamma=2$ by mapping onto free fermions
\cite{Jancovici81}.
The 2D TCP undergoes the collapse of positive-negative pairs of pointlike 
charges at coupling $\Gamma=2$, which corresponds to the exactly solvable 
free-fermion point of the equivalent Thirring model \cite{Cornu87,Gaudin85}. 
For reviews about exactly solvable 2D Coulomb systems with various geometries
of confining domains, see Refs. \cite{Forrester98,Jancovici92}.

At arbitrary temperature of the conducting regime, the long-range tail of 
the Coulomb potential induces screening.
Due to the screening effect, the bulk particle correlations exhibit 
short-range (exponential or even Gaussian) decay which indicates a
non-criticality.
On the other hand, the same screening phenomenon causes that the induced 
electrical-field correlations are long-ranged \cite{Jancovici95,Lebowitz84}.
As a result, the free energy (or the grand potential) of any 
Coulomb system exhibits a universal finite-size correction of type 
(\ref{finite-size}).
Since the logarithmic Coulomb potential is the inverse of the Laplacian
operator $\Delta$, in the functional sense it is associated with the Gaussian
Hamiltonian (\ref{GaussH}) rewritten as 
$-\int_D {\rm d}^2r\, \phi \Delta\phi/(4\pi)$.
Although the conformal anomaly number of Coulomb systems is expected to be 
related to the Gaussian one, there is a change of sign, namely 
\begin{equation}
c=-1 .
\end{equation}

The explicit checks of the universal finite-size behavior with $c=-1$ 
were done, for both OCP and TCP, at the exactly solvable coupling $\Gamma=2$.
The studied cases involve Coulomb gases with periodic boundary conditions
\cite{Forrester91}, confined to a domain by plain hard walls 
\cite{Jancovici94}, by ideal-conductor boundaries \cite{Jancovici96} and
by ideal-dielectric walls \cite{Jancovici01,Tellez01}.
For Coulomb systems living on a surface of the sphere, a direct derivation
of the universal finite-size correction was performed for any coupling
by combining stereographic projection of the sphere onto an infinite plane
with linear response theory (TCP, Ref. \cite{Jancovici00a}) or with
density functional approach (OCP, Ref. \cite{Jancovici00c}).
The prefactor to the universal correction term was related to the second
moment of the short-range part of the planar direct correlation function.
Using a renormalized Mayer expansion \cite{Deutsch74,Friedman62} and
observing a cancellation property of specific families of renormalized diagrams,
this second moment was evaluated for both TCP \cite{Jancovici00b}
and OCP \cite{Kalinay00}. 
Another explicit derivation of the universal logarithmic term at an arbitrary 
coupling constant was made for the TCP in a disk geometry \cite{Samaj02}. 
All obtained results confirm the conformal prediction (\ref{finite-size}) 
for a critical system as we had $c=-1$.

All that has been said holds for neutral Coulomb systems.
Recently, Ferrero and T\'ellez \cite{Ferrero14} studied the 2D TCP of $\pm e$ 
charges at the exactly solvable collapse $\Gamma=2$ point, confined to 
a disk of radius $R$.
They fixed a ``guest'' hard-core impurity of charge $Q e$ at 
the disk origin and found that the large-$R$ expansion for 
the grand potential of type (\ref{finite-size}) applies if 
the conformal anomaly number is modified to
\begin{equation} \label{cQGamma2}
c(Q,\Gamma=2) = -1 + 6 Q^2 ;
\end{equation}
presence of the impurity has no effect on bulk and boundary terms.
It has been noted in the same reference that the same modification of 
the anomaly number occurs when the minimal free boson conformal field theory 
\cite{Dotsenko04,Ginsparg89}, equivalent formally to the 2D TCP, 
is deformed by spreading out a charge $Q e$ at infinity, creating in 
this way a non-neutral system.
The equivalence of the two systems, one with a charge fixed at the disk 
origin and one with the same charge spread out at infinity, can be explained
by noting that a charge fixed at the disk origin is screened at microscopic 
distances by counterions from the TCP and thus the universal large-$R$ 
correction looks like the one for a non-neutral system with 
the net charge $Q e$.
Notice that the definition of $c$ in Refs. \cite{Dotsenko04,Ferrero14} has
opposite sign with respect to the standard notation used in this paper. 
At an arbitrary $\Gamma$, conformal field theory \cite{Dotsenko04,Ginsparg89}
yields
\begin{equation} \label{cQGamma}
c(Q,\Gamma) = -1 + 3 \Gamma Q^2 .
\end{equation}
With this conformal number, the logarithmic correction in (\ref{finite-size}) 
is no longer universal, its prefactor depends on both the coupling constant 
and the excess charge.

In this paper, we concentrate on the non-neutral 2D OCP confined to the disk.
We perform a microscopic derivation of the finite-size expansion of
the free energy and find that it is of type (\ref{finite-size}) 
with the same anomaly number (\ref{cQGamma}) as found for the 2D TCP
in the context of conformal field theories.
We would like to note that in contrast to the previous studies 
\cite{Jancovici03,Levesque00,Martin80} of charge fluctuations in finite 
Coulomb systems in a much larger overall neutral system or with particle 
reservoir put at infinity, here we consider non-neutral systems 
within the constrained canonical ensemble with the fixed net 
charge and look for the dependence of finite-size corrections of 
the free energy on this excess charge.  

The paper is organized as follows.
In Sect. 2, we derive the total energy of mobile charged particles and
the fixed background in presence of non-neutrality.
Section 3 summarizas the general mapping of the 2D OCP onto the theory
of anticommuting variables on a one-dimensional chain.
The exactly solvable free-fermion case is the subject of Sect. 4.
The free-fermion result is generalized to an arbitrary coupling in Sect. 5.
In Sect. 6, we give a short recapitulation and some concluding remarks. 

\renewcommand{\theequation}{2.\arabic{equation}}
\setcounter{equation}{0}

\section{Derivation of Boltzmann factor} \label{Sect.2}
Let us consider the 2D OCP inside the disk $D$ of radius $R$.
There are $N$ pointlike particles of charge $e$ immersed in a uniform
background charge density $\rho_b = -e n_b$, where the background density 
$n_b$ is given by
\begin{equation}
\pi R^2 n_b = N - Q , \qquad \mbox{finite $Q<N$}. 
\end{equation}
The net charge of the system is thus $Q e$ and the neutral case corresponds
to $Q=0$. 
The particle density $n$ is given by
\begin{equation}
\pi R^2 n = N .
\end{equation} 
As
\begin{equation} \label{ratio}
\frac{n_b}{n} = 1 - \frac{Q}{N} ,
\end{equation}
the particle and background densities coincide in the thermodynamic 
limit $N\to\infty$. 

The potential induced by the background charge density reads as
\begin{equation}
V({\bf r}) = - \int_D {\rm d}^2r'\, \rho_b 
\ln \left\vert \frac{{\bf r}-{\bf r}'}{L} \right\vert .
\end{equation}
This potential satisfies the Poisson equation with circular symmetry
\begin{equation} \label{polarPoisson}
\Delta V({\bf r}) \equiv \frac{1}{r} \frac{\partial}{\partial r}
\left[ r \frac{\partial}{\partial r} V(r) \right] = 2\pi e n_b .
\end{equation}
The potential at the disk boundary is given by the total background charge 
placed at the origin, i.e., $V(R)= e n_b (\pi R^2) \ln(R/L)$.
The solution of Eq. (\ref{polarPoisson}), supplemented by this boundary
condition, reads
\begin{equation}
V(r) = e n_b \pi \left[ \frac{r^2}{2} + 
R^2 \ln\left( \frac{R}{L} \right) - \frac{R^2}{2} \right] .
\end{equation}

To derive the total Boltzmann factor, we proceed in close analogy
with neutral systems, see e.g. Refs. \cite{DiFrancesco94,Forrester98,Sari76}.
The total Coulomb interaction energy of the particle-background system
consists of three parts:
\begin{itemize}
\item the particle-particle interaction
\begin{eqnarray}
U_1 & = & - e^2 \sum_{1\le j<k\le N} 
\ln \frac{\vert {\bf r}_j-{\bf r}_k\vert}{L} \nonumber \\
& = & \frac{e^2}{2} N(N-1) \ln L - e^2 \sum_{1\le j<k\le N} 
\ln \left\vert {\bf r}_j-{\bf r}_k \right\vert ,
\end{eqnarray}
\item the particle-background interaction
\begin{eqnarray}
U_2 & = & e \sum_{j=1}^N V({\bf r}_j) \nonumber \\
& = & \frac{e^2 n_b \pi}{2} \sum_{j=1}^N r_j^2 + e^2 N(N-Q) \left[ 
\ln\left( \frac{R}{L} \right) - \frac{1}{2} \right] ,
\end{eqnarray}
\item the background-background interaction
\begin{eqnarray}
U_3 & = & - \frac{1}{2} \int_D {\rm d}^2r \int_D {\rm d}^2r'\, 
\rho_b^2 \ln \left\vert \frac{{\bf r}-{\bf r}'}{L} \right\vert \nonumber \\
& = & - \frac{e^2}{2} (N-Q)^2 \left[ 
\ln\left( \frac{R}{L} \right) - \frac{1}{4} \right] .
\end{eqnarray}
\end{itemize}
With the definition of the coupling constant $\Gamma=\beta e^2 \equiv 2\gamma$,
the corresponding Boltzmann factor reads as
\begin{eqnarray}
{\rm e}^{-\beta(U_1+U_2+U_3)} & = & \exp\Bigg[ \gamma Q^2 
\ln\left( \frac{R}{L}\right) + \gamma N\ln L - \gamma N^2\ln R + \frac{3}{4} 
\gamma N^2 \nonumber \\ & & 
- \frac{1}{2} \gamma N Q - \frac{1}{4} \gamma Q^2 \Bigg] 
\prod_{j=1}^N {\rm e}^{-\pi\gamma n_b r_j^2} 
\prod_{1\le j<k\le N} \left\vert {\bf r}_j-{\bf r}_k \right\vert^{2\gamma} . 
\phantom{aaaa}
\end{eqnarray}
From the expression for the canonical partition function
\begin{equation} \label{partition}
Z_{\gamma}(N,Q) = \frac{1}{N!} \int_D \prod_{j=1}^N 
\frac{{\rm d}^2 r_j}{\lambda^2}\, {\rm e}^{-\beta(U_1+U_2+U_3)} ,
\end{equation}
the free energy $F_{\gamma}(N,Q)$ is given by
$-\beta F_{\gamma}(N,Q) = \ln Z_{\gamma}(N,Q)$, i.e.,
\begin{eqnarray}
-\beta F_{\gamma}(N,Q) & = & \gamma Q^2 \ln\left( \frac{R}{L}\right) 
+ \gamma N\ln L - \gamma N^2\ln R + \frac{3}{4} \gamma N^2
\nonumber \\ & & - \frac{1}{2} \gamma N Q 
- \frac{1}{4} \gamma Q^2 - 2N\ln\lambda + \ln I_{\gamma}(N,Q) . \label{free}
\end{eqnarray}
Here,
\begin{equation} \label{conf}
I_{\gamma}(N,Q) = \frac{1}{N!} \int_D \prod_{j=1}^N 
\left[ {\rm d}^2 r_j\, w({\bf r}_j) \right] 
\prod_{1\le j<k\le N} \vert {\bf r}_j-{\bf r}_k\vert^{2\gamma} 
\end{equation}
is the configuration integral with the circularly symmetric one-body 
Boltzmann factor 
\begin{equation} \label{Boltzmann}
w({\bf r}) = \exp\left( -\gamma \pi n_b r^2 \right) .
\end{equation}

\renewcommand{\theequation}{3.\arabic{equation}}
\setcounter{equation}{0}

\section{Mapping onto 1D fermions} \label{Sect.3}
There exist two basic approaches how to treat the configuration integral 
(\ref{conf}).
The first one is based on the expansion of the power of van der monde
determinants in the basis of Jack polynomials 
\cite{Salazar16,Tellez99,Tellez12}.
The second one is based on the mapping onto a 1D lattice fermion system;
this formalism has been introduced in Ref. \cite{Samaj95} and developed 
further in Refs. \cite{Samaj00,Samaj04,Samaj11,Samaj14,Samaj15,Samaj16}.
The relation between the two methods was established in Ref. \cite{Grimaldo15}.
Here, we apply the mapping onto 1D fermions.

For $\gamma$ a positive integer, the configuration integral (\ref{conf}) 
can be expressed in terms of two sets of anticommuting variables 
$\{ \xi_j^{(\alpha)},\psi_j^{(\alpha)} \}$ each with $\gamma$ components 
$(\alpha=1,\ldots,\gamma)$, defined on a discrete chain of $N$ sites 
$j=0,1,\ldots,N-1$, as follows
\begin{equation} \label{anticonf}
I_{\gamma} = \int {\cal D}\psi {\cal D}\xi\, {\rm e}^{{\cal S}(\xi,\psi)} , 
\qquad {\cal S}(\xi,\psi) = \sum_{j=0}^{\gamma(N-1)} w_j \Xi_j \Psi_j .
\end{equation}
Here, ${\cal D}\psi {\cal D}\xi \equiv \prod_{j=0}^{N-1} {\rm d}\psi_j^{(\gamma)}
\cdots {\rm d}\psi_j^{(1)} {\rm d}\xi_j^{(\gamma)} \cdots {\rm d}\xi_j^{(1)}$
and the action ${\cal S}(\xi,\psi)$ involves pair interactions of composite
operators
\begin{equation} \label{composite}
\Xi_j = \sum_{j_1,\ldots,j_{\gamma}=0\atop (j_1+\cdots+j_{\gamma}=j)}^{N-1}
\xi_{j_1}^{(1)} \cdots \xi_{j_{\gamma}}^{(\gamma)} , \qquad
\Psi_j = \sum_{j_1,\ldots,j_{\gamma}=0\atop (j_1+\cdots+j_{\gamma}=j)}^{N-1}
\psi_{j_1}^{(1)} \cdots \psi_{j_{\gamma}}^{(\gamma)} ,
\end{equation} 
i.e. the products of $\gamma$ anticommuting variables from a given set
with the fixed sum of site indices.
The elements of the diagonal interaction matrix are given by
\begin{equation} \label{wgeneral}
w_j = \int_D {\rm d}^2 r\, r^{2j} w(r) , \qquad j=0,1,\ldots,\gamma(N-1) .
\end{equation}
For the present one-body Boltzmann factor (\ref{Boltzmann}), the interaction 
strengths can be rewritten as follows
\begin{equation} \label{wour}
w_j = \frac{1}{\gamma n_b} \frac{1}{(\gamma\pi n_b)^j} \tilde{w}_j , \qquad
\tilde{w}_j = \int_0^{\gamma(N-Q)} {\rm d}t\, t^j {\rm e}^{-t} . 
\end{equation}

The one-body density of particles at point ${\bf r}\in D$ is defined by
\begin{equation}
n({\bf r}) = \langle \hat{n}({\bf r}) \rangle , \qquad
\hat{n}({\bf r}) = \sum_{j=1}^N \delta({\bf r}-{\bf r}_j) ,
\end{equation}
where $\hat{n}({\bf r})$ is the microscopic density of particles at point
${\bf r}$ and $\langle \cdots \rangle$ denotes the statistical average over
canonical ensemble.
The one-body density can be obtained from the configuration integral 
(\ref{conf}) in the standard way as the functional derivative:
\begin{equation}
n({\bf r}) = w({\bf r}) \frac{1}{I_{\gamma}} 
\frac{\delta I_{\gamma}}{\delta w({\bf r})} .
\end{equation}
Within the formalism of anticommuting variables, the circularly symmetric
one-body density is expressible explicitly as
\begin{equation} \label{density}
n(r) = w(r) \sum_{j=0}^{\gamma(N-1)} 
\langle \Xi_j \Psi_j \rangle r^{2j} , 
\end{equation}
where 
\begin{equation}
\langle \cdots\rangle \equiv \frac{1}{I_{\gamma}}
\int {\cal D}\psi {\cal D}\xi\, {\rm e}^S \cdots
\end{equation} 
denotes averaging over anticommuting variables. 

\renewcommand{\theequation}{4.\arabic{equation}}
\setcounter{equation}{0}

\section{The free-fermion point $\Gamma=2$} \label{Sect.4}
For $\Gamma=2$ $(\gamma=1)$, the composite operators (\ref{composite})
are the ordinary anticommuting variables. 
The diagonalized action in (\ref{anticonf}) then implies
\begin{equation}
I_1(N,Q) = \prod_{j=0}^{N-1} w_j .
\end{equation}
Since the correlators
\begin{equation}
\langle \Xi_j \Psi_j \rangle = \frac{1}{w_j} , \qquad j=0,1,\ldots,N-1, 
\end{equation}
the particle density (\ref{density}) takes the form
\begin{equation}
n(N,Q;r) = {\rm e}^{-\pi n_b r^2} \sum_{j=0}^{N-1} \frac{r^{2j}}{w_j} .
\end{equation}

According to (\ref{wour}), the interaction strengths are expressible as
\begin{equation} \label{wour1}
w_j = \frac{1}{n_b} \frac{1}{(\pi n_b)^j} \tilde{w}_j , \qquad
\tilde{w}_j = \int_0^{N-Q} {\rm d}t\, t^j {\rm e}^{-t} . 
\end{equation}
Since
\begin{equation}
\ln I_1(N,Q) = - N \ln n_b - \frac{1}{2} N(N-1) \ln(\pi n_b)
+ \sum_{j=0}^{N-1} \ln \tilde{w}_j ,
\end{equation}
the expression (\ref{free}) for the free energy can be rewritten as
\begin{eqnarray}
-\beta F_1(N,Q) & = & Q^2 \ln\left( \frac{R}{L} \right)
+ \frac{1}{2} N \ln\left( \pi n_b L^2\right) 
- \frac{1}{2} N^2 \ln(N-Q) \nonumber \\ & & 
+ \frac{3}{4} N^2 - \frac{1}{2} N Q - \frac{1}{4} Q^2 
- N \ln\left( n_b\lambda^2\right) 
+ \sum_{j=0}^{N-1} \ln \tilde{w}_j . \label{freenew}
\end{eqnarray}

Now we aim at estimating the crucial sum $\sum_{j=0}^{N-1} \ln \tilde{w}_j$.
Since it holds
\begin{equation}
\tilde{w}_j = \int_0^N {\rm d}t\, t^j {\rm e}^{-t} -
{\rm e}^{-N}\int_0^Q {\rm d}s\, (N-s)^j {\rm e}^s ,  
\end{equation}
we get the important relation
\begin{equation}
\frac{\partial \tilde{w}_j}{\partial Q} = - {\rm e}^{-(N-Q)} (N-Q)^j .
\end{equation}
Consequently,
\begin{eqnarray}
\frac{\partial}{\partial Q} \sum_{j=0}^{N-1} \ln \tilde{w}_j
& = & - {\rm e}^{-(N-Q)} \sum_{j=0}^{N-1} \frac{1}{\tilde{w}_j} (N-Q)^j 
\nonumber \\
& = & - \frac{1}{n_b} {\rm e}^{-(N-Q)} \sum_{j=0}^{N-1} \frac{1}{w_j} 
\left( \frac{N-Q}{\pi n_b} \right)^j \nonumber \\
& = & - \frac{1}{n_b} {\rm e}^{-\pi n_b R^2} 
\sum_{j=0}^{N-1} \frac{R^{2j}}{\tilde{w}_j} = - \frac{n(N,Q;R)}{n_b} ,
\end{eqnarray}
where $n(N,Q;R)$ is the particle number density at $r=R$, i.e., 
at the wall contact.
We can write
\begin{equation} \label{suma}
\sum_{j=0}^{N-1} \ln \tilde{w}_j = \sum_{j=0}^{N-1} \ln j! + \sum_{j=0}^{N-1} 
\ln \left( \frac{1}{j!} \int_0^N {\rm d}t\, t^j {\rm e}^{-t} \right) 
- \int_0^{Q} {\rm d}Q'\, \frac{n(N,Q';R)}{n_b} .
\end{equation}
This formula completes the determination of the free energy (\ref{freenew}). 

Let us now consider the large-$N$ limit when the background density is 
related to the particle density via $n_b=n+O(1/N)$.
We start with the sum representation (\ref{suma}).
With the aid of formulas presented in Appendix A of Ref. \cite{DiFrancesco94}, 
we find that  
\begin{equation}
\sum_{j=0}^{N-1} \ln j! = \frac{1}{2} N^2 \ln N - \frac{3}{4} N^2
+ N \frac{\ln(2\pi)}{2} - \frac{1}{12} \ln N + O(1) .
\end{equation}
Using the asymptotic formula \cite{Erdelyi}
\begin{equation}
\frac{1}{j!} \int_0^N {\rm d}t\, t^j {\rm e}^{-t}
= \frac{1}{2} \left[ 1 + \Phi\left( \frac{N-n}{\sqrt{2N}} \right) \right]
+ O\left( \frac{1}{\sqrt{N}}\right) 
\end{equation}
with
\begin{equation}
\Phi(u) = \frac{2}{\sqrt{\pi}} \int_0^u {\rm d}v\, {\rm e}^{-v^2}
\end{equation}
being the error function, we obtain \cite{Jancovici94}
\begin{equation}
\sum_{j=0}^{N-1} \ln \left( \frac{1}{j!} 
\int_0^N {\rm d}t\, t^j {\rm e}^{-t} \right) =
\sqrt{2N} \int_0^{\infty} {\rm d}v\, \ln\left[ \frac{1+\Phi(v)}{2} \right]
+ O(1) .
\end{equation}
Finally, the integral
\begin{equation}
\int_0^{Q} {\rm d}Q'\, \frac{n(N,Q';R)}{n_b}
\end{equation}
depends on the statistical average $n(N,Q';R)$ which is finite for any $N$, 
and therefore in the limit $N\to\infty$ contributes to $O(1)$. 
It is natural to suppose that in the limit $N\to\infty$ the statistical 
averages do not depend on the finite charge imbalance $Q'e$ within an infinite 
domain and so the particle density at the disk border $n(N,Q';R)$ tends to 
its half-space ``wall contact'' value $n_{\rm wall} = n\ln 2$ 
\cite{Jancovici82}, independent of $Q'$. 
Thus we have 
\begin{equation}
\lim_{N\to\infty} \int_0^{Q} {\rm d}Q'\, \frac{n(N,Q';R)}{n_b} = Q \ln 2
\end{equation}
which is the term of order $O(1)$.
To summarize,
\begin{eqnarray}
\sum_{j=0}^{N-1} \ln \tilde{w}_j & = & 
\frac{1}{2} N^2 \ln N - \frac{3}{4} N^2
+ N \frac{\ln(2\pi)}{2} - \frac{1}{12} \ln N \nonumber \\ & & 
+ \sqrt{2N} \int_0^{\infty} {\rm d}v\, \ln\left[ \frac{1+\Phi(v)}{2} \right]
+ O(1) .
\end{eqnarray}
Inserting this large-$N$ formula into (\ref{freenew}) and taking into account
that
\begin{equation}
- \frac{1}{2} N^2 \ln(N-Q) - \frac{1}{2}NQ = - \frac{1}{2} N^2\ln N + O(1) ,
\end{equation}
we arrive at the finite-size expansion of type (\ref{finite-size}) with
\begin{eqnarray}
\beta f & = & n \left[ \ln(n\lambda^2) - \frac{1}{2} \ln(\pi n L^2)
- \frac{1}{2} \ln(2\pi) \right] , \nonumber \\
\beta\sigma & = & - \sqrt{\frac{n}{2\pi}}
\int_0^{\infty} {\rm d}v\, \ln\left[ \frac{1+\Phi(v)}{2} \right] , \nonumber \\ 
c(Q,\Gamma=2) & = & -1 + 6 Q^2 .
\end{eqnarray}
We see that for the coupling $\Gamma=2$ the central charge depends on $Q$ 
in the same way as for the 2D TCP at the same coupling constant $\Gamma=2$, 
see Eq. (\ref{cQGamma2}).

\renewcommand{\theequation}{5.\arabic{equation}}
\setcounter{equation}{0}

\section{Arbitrary $\Gamma=2\gamma$ with integer $\gamma$} \label{Sect.5}
For an arbitrary integer $\gamma$, there exist certain analogies with 
the exactly solvable $\gamma=1$ case.

In the configuration integral (\ref{anticonf}), we perform the unitary 
transformation of anticommuting variables, say $\xi$'s, 
\begin{equation} \label{transformation}
\xi_j \to \mu^{1/\gamma} \lambda^j \tilde{\xi}_j , \qquad
\Xi_j \to \mu \lambda^j \tilde{\Xi}_j , 
\end{equation}
which keeps the composite form of $\Xi$-operators.
Here, $\mu$ and $\lambda$ are free as-yet unspecified parameters.
Taking into account the corresponding Jacobian, this transformation modifies 
$I_{\gamma}(N,Q)$ to the form 
\begin{equation}
I_{\gamma}(N,Q) = \mu^{-N} \lambda^{-\gamma N(N-1)/2} \tilde{I}_{\gamma}(N,Q) , 
\end{equation}
where
\begin{equation}
\tilde{I}_{\gamma}(N,Q) = \int {\cal D}\psi {\cal D}\tilde{\xi}\, 
\exp\left(\sum_{j=0}^{\gamma(N-1)} \mu \lambda^j w_j \tilde{\Xi}_j \Psi_j\right) .
\end{equation}
Choosing
\begin{equation}
\mu = \gamma n_b , \qquad \lambda = \gamma \pi n_b 
\end{equation}
we eliminate the $Q$-dependent prefactor in (\ref{wour}) and in this way
pass from $w_j$ to $\tilde{w}_j$.
We obtain that
\begin{equation} \label{Igamma}
\ln I_{\gamma}(N,Q) = - N \ln(\gamma n_b) - \frac{\gamma}{2} N(N-1) 
\ln(\gamma \pi n_b) + \ln \tilde{I}_{\gamma}(N,Q) , 
\end{equation}
where
\begin{equation}
\tilde{I}_{\gamma}(N,Q) = \int {\cal D}\psi {\cal D}\tilde{\xi}\, 
\exp\left(\sum_{j=0}^{\gamma(N-1)} \tilde{w}_j \tilde{\Xi}_j \Psi_j\right) .
\end{equation}

To obtain a convenient representation of $\ln\tilde{I}_{\gamma}(N,Q)$,
we differentiate this quantity with respect to $Q$:
\begin{equation} \label{tildeI}
\frac{\partial}{\partial Q} \ln\tilde{I}_{\gamma}(N,Q)
= \sum_{j=0}^{\gamma(N-1)} \langle \tilde{\Xi}_j \Psi_j \rangle
\frac{\partial\tilde{w}_j}{\partial Q} . 
\end{equation}
Since
\begin{equation}
\tilde{w}_j = \int_0^{\gamma N} {\rm d}t\, t^j {\rm e}^{-t} -
\gamma {\rm e}^{-\gamma N}\int_0^Q {\rm d}s\, [\gamma(N-s)]^j {\rm e}^{\gamma s} ,  
\end{equation}
we get
\begin{equation}
\frac{\partial \tilde{w}_j}{\partial Q} = - \gamma {\rm e}^{-\gamma(N-Q)} 
[\gamma(N-Q)]^j .
\end{equation}
As concerns the correlators $\langle\tilde{\Xi}_j\Psi_j\rangle$, we return
back to the original anticommuting $\xi$'s variables by applying the inverse
to the transformation (\ref{transformation}).
The Jacobians of the numerator and the denominator cancel with one another
and we obtain
\begin{equation}
\langle \tilde{\Xi}_j \Psi_j \rangle = \frac{1}{\mu \lambda^j}
\langle \Xi_j \Psi_j \rangle = \frac{1}{\gamma n_b} \frac{1}{(\gamma\pi n_b)^j}
\langle \Xi_j \Psi_j \rangle .
\end{equation}
Consequently, the relation (\ref{tildeI}) takes the form
\begin{eqnarray}
\frac{\partial}{\partial Q} \ln\tilde{I}_{\gamma}(N,Q)
& = & - \frac{1}{n_b} {\rm e}^{-\gamma(N-Q)}
\sum_{j=0}^{\gamma(N-1)} \langle \Xi_j \Psi_j \rangle
\left( \frac{N-Q}{\pi n_b} \right)^j \nonumber \\
& = & - \frac{1}{n_b} {\rm e}^{-\gamma\pi n_b R^2}
\sum_{j=0}^{\gamma(N-1)} \langle \Xi_j \Psi_j \rangle R^{2j}
= - \frac{n(N,Q;R)}{n_b} . \phantom{aaa} 
\end{eqnarray}
As the result
\begin{equation} \label{final}
\ln\tilde{I}_{\gamma}(N,Q) = \ln\tilde{I}_{\gamma}(N,Q=0) 
- \int_0^Q {\rm d}Q'\, \frac{n(N,Q';R)}{n_b} . 
\end{equation}

Putting together Eqs. (\ref{free}), (\ref{Igamma}) and (\ref{final}),
the free energy is given by
\begin{eqnarray}
\beta F_{\gamma}(N,Q) & = & - \gamma Q^2 \ln\left( \frac{R}{L}\right) 
- \frac{1}{2} \gamma N\ln\left( \gamma\pi n_b L^2\right) 
+ \frac{1}{2} \gamma N^2\ln\left[ \gamma(N-Q)\right]  \nonumber \\ & &  
- \frac{3}{4} \gamma N^2 + \frac{1}{2} \gamma N Q
+ \frac{1}{4} \gamma Q^2 + N\ln\left( \gamma n_b \lambda^2 \right) 
\nonumber \\ & &
- \ln \tilde{I}_{\gamma}(N,Q=0) + \int_0^Q {\rm d}Q'\, \frac{n(N,Q';R)}{n_b} . 
\end{eqnarray}
As before, the mean particle density at the disk border $n(N,Q';R)$ 
is finite for any $N$ and in the large-$N$ limit we have
\begin{equation}
\lim_{N\to\infty} \int_0^{Q} {\rm d}Q'\, \frac{n(N,Q';R)}{n_b} 
= Q \frac{n_{\rm wall}}{n} .
\end{equation}
In contrast to the $\gamma=1$ case, the explicit value of $n_{\rm wall}/n$
is not known, but the important information is that
this term is of order $O(1)$.
Consequently,
\begin{equation}
\beta F_{\gamma}(N,Q) = \beta F_{\gamma}(N,Q=0) - \frac{\Gamma}{2} Q^2 
\ln\left( \frac{R}{L}\right) + O(1) . 
\end{equation}
Considering for the neutral $\beta F_{\gamma}(N,Q=0)$ the anticipated critical 
finite-size expansion (\ref{finite-size}) with $c=-1$, the conformal anomaly 
number is equal to the one predicted by conformal field theory in
Eq. (\ref{cQGamma}).

Although the proof was made for the 2D OCP with $\gamma$ a positive integer, 
it is reasonable to extend its validity to all real coupling values 
$\Gamma>0$, within the fluid regime. 

\renewcommand{\theequation}{6.\arabic{equation}}
\setcounter{equation}{0}

\section{Conclusion} \label{Sect.6}
This work was motivated by the recent paper of Ferrero and T\'ellez 
\cite{Ferrero14} about the finite-size expansion of the grand potential
for the 2D TCP at the coupling $\Gamma=2$, in the presence of the impurity.
The modification of the universal logarithmic term to the non-universal one,
dependent on the charge $Q e$ of impurity, was there in agreement with
the prediction (\ref{cQGamma2}) of non-neutral conformal field theories.

In this paper, we tested the prediction (\ref{cQGamma2}) of conformal 
field theory on the non-neutral 2D OCP.
At the free-fermion coupling $\Gamma=2$ (Sect. 4), we reproduce  
the result (\ref{cQGamma2}). 
Using the mapping of the 2D OCP onto the anticommuting field theory 
formulated on the chain, we were able to extend the finite-size 
analysis to an arbitrary coupling constant $\Gamma = 2*{\rm integer}$.
This is one of rare occasions when the exact results are obtained
for a series of pair-integer $\Gamma$ values. 
The non-universal prediction of conformal field theory (\ref{cQGamma}) 
is confirmed as well.

It would be interesting to generalize our results to the 2D TCP or
even to Coulomb fluids with an arbitrary charge composition.
For such systems, we miss techniques analogous to that
for 2D OCP, but a phenomenological approach might reveal 
the general form of the finite-size logarithmic term.

This work might be a further motivation for specialists to use Coulomb fluids 
as practical realizations of conformal field theories
to test their predictions. 

\begin{acknowledgements}
The support received from Grant VEGA No. 2/0015/15 is acknowledged. 
\end{acknowledgements}


\begin{thebibliography}{10}

\bibitem{Affleck86} Affleck, I.:
Universal term in the free energy at a critical point and 
the conformal anomaly.
Phys. Rev. Lett. {\bf 56}, 746--748 (1986)

\bibitem{Blote86} Bl\"ote, H.W.J., Cardy, J.L., Nightingale, M.P.:
Conformal invariance, the central charge, and universal finite-size 
amplitudes at criticality.
Phys. Rev. Lett. {\bf 56}, 742--745 (1986)

\bibitem{Cardy88a} Cardy, J.L., Peschel, I.:
Finite-size dependence of the free energy in two-dimensional critical systems.
Nucl. Phys. B {\bf300}, 377--392 (1988)

\bibitem{Cardy88b} Cardy, J.L.:
Conformal invariance and statistical mechanics.
In: Br\'ezin, E, Zinn-Justin, J. (eds.), Fields, Strings and Critical 
Phenomena, Les Houches 1988, Session XLIX, North-Holland, Amsterdam (1990) 

\bibitem{Cornu87} Cornu, F., Jancovici, B.:
On the two-dimensional Coulomb gas.
J. Stat. Phys. {\bf 49}, 33--56 (1987)

\bibitem{DiFrancesco94} Di Francesco, P., Gaudin, M., Itzykson, C., Lesage, F.:
Laughlin's wave functions, Coulomb gases and expansions of the discriminant.
Int. J. Mod. Phys. A {\bf 9}, 4257--4351 (1994)

\bibitem{Deutsch74} Deutsch, C., Lavaud, M.:
Equilibrium properties of a two-dimensional Coulomb gas.
Phys. Rev. A {\bf 9}, 2598--2616 (1974)

\bibitem{Dotsenko04} Dotsenko, V.S.:
Serie de cours sur la th\'eorie conform.
Universit\'e de Paris VI-VII (2004)

\bibitem{Erdelyi} Erd\'elyi, A.:
Higher Transcendental Functions. McGraw-Hill, New York (1953)

\bibitem{Ferrero14} Ferrero, A., T\'ellez, G.:
Screening of an electrically charged particle in a two-dimensional
two-component plasma at $\Gamma=2$.
J. Stat. Mech., P11021 (2014) 

\bibitem{Forrester91} Forrester, P.J.:
Finite-size corrections to the free energy of Coulomb systems
with a periodic boundary condition.
J. Stat. Phys. {\bf 63}, 491--504 (1991)

\bibitem{Forrester98} Forrester, P.J.:
Exact results for two-dimensional Coulomb systems.
Phys. Rep. {\bf 301}, 235--270 (1998)

\bibitem{Friedman62} Friedman, H.L.:
Ionic solution theory. Interscience, New York (1962)
 
\bibitem{Gaudin85} Gaudin, M.:
Critical isotherm of a lattice plasma.
J. Phys. France {\bf 46} 1027--1042 (1985)

\bibitem{Ginsparg89} Ginsparg, P.:
Applied conformal field theory.
In: Br\'ezin, E, Zinn-Justin, J. (eds.), Fields, Strings and Critical 
Phenomena, Les Houches 1988, Session XLIX, North-Holland (1990) 

\bibitem{Grimaldo15} Grimaldo, J.A.M., Tell\'ez, G.: 
Relations among two methods for computing the partition function
of the two-dimensional one-component plasma.
J. Stat. Phys. {\bf 160}, 4--28 (2015)

\bibitem{Jancovici81} Jancovici, B.:
Exact results for the two-dimensional one-component plasma.
Phys. Rev. Lett. {\bf 46}, 386--388 (1981)

\bibitem{Jancovici82} Jancovici, B.:
Classical Coulomb systems near a plane wall. I.
J. Stat. Phys. {\bf 28}, 43--65 (1982)

\bibitem{Jancovici92} Jancovici, B.:
Inhomogeneous two-dimensional plasmas.
In: Henderson. D. (ed.) Inhomogeneous Fluids, pp. 201--237, Dekker, 
New York (1992)

\bibitem{Jancovici94} Jancovici B., Manificat, G., Pisani, C.:
Coulomb systems seen as critical systems:
Finite-size effects in two dimensions.
J. Stat. Phys. {\bf 76}, 307--329 (1994)

\bibitem{Jancovici95} Jancovici, B.:
Classical Coulomb systems: Screening and correlations revisited.
J. Stat. Phys. {\bf 80}, 445--459 (1995)

\bibitem{Jancovici96} Jancovici, B., Tell\'ez, G.:
Coulomb systems seen as critical systems: Ideal conductor boundaries.
J. Stat. Phys. {\bf 82}, 609--632 (1996)

\bibitem{Jancovici00a} Jancovici, B.:
A sum rule for the two-dimensional two-component plasma.
J. Stat. Phys. {\bf 100}, 201--207 (2000)

\bibitem{Jancovici00b} Jancovici, B., Kalinay, P., \v{S}amaj, L.:
Another derivation of a sum rule for the two-dimensional two-component plasma.
Physica A {\bf 279}, 260--267 (2000)

\bibitem{Jancovici00c} Jancovici, B., Trizac, E.:
Universal free energy correction for the two-dimensional one-component plasma.
Physica A {\bf 284}, 241--245 (2000)

\bibitem{Jancovici01} Jancovici, B., \v{S}amaj, L.:
Coulomb systems with ideal dielectric boundaries:
Free fermion point and universality.
J. Stat. Phys. {\bf 104}, 753--775 (2001)

\bibitem{Jancovici03} Jancovici:
Charge fluctuations in finite Coulomb systems.
J. Stat. Phys. {\bf 110}, 879--902 (2003)

\bibitem{Kalinay00} Kalinay, P., Marko\v{s}, P., \v{S}amaj, L., 
Trav\v{e}nec, I.:
The sixth-moment sum rule for the pair correlations of the two-dimensional
one-component plasma: Exact result.
J. Stat. Phys. {\bf 98}, 639--666 (2000)

\bibitem{Lebowitz84} Lebowitz, J.L., Martin, Ph.A.:
On potential and field fluctuations in classical charged systems.
J. Stat. Phys. {\bf 34}, 287--311 (1984).

\bibitem{Levesque00} Levesque, D., Weis, J.-J., Lebowitz, J.L.:
Charge fluctuations in the two-dimensional one-component plasma.
J. Stat. Phys. {\bf 100}, 209--222 (2000).

\bibitem{Martin80} Martin, Ph.A., Yalcin, T.:
The charge fluctuations in classical Coulomb systems.
J. Stat. Phys. {\bf 22}, 435--463 (1980)

\bibitem{Salazar16} Salazar, R., Tell\'ez, G.: 
Exact energy computation of the one component plasma on a sphere for
even values of the coupling parameter,
J. Stat. Phys. {\bf 164}, 969--999 (2016)

\bibitem{Samaj95} \v{S}amaj, L., Percus, J.K.:
A functional relation among the pair correlations of the two-dimensional
one-component plasma.
J. Stat. Phys. {\bf 80}, 811--824 (1995)

\bibitem{Samaj00} \v{S}amaj, L.:
Microscopic calculation of the dielectric susceptibility tensor for 
Coulomb fluids.
J. Stat. Phys. {\bf 100}, 949--967 (2000)

\bibitem{Samaj02} \v{S}amaj, L., Jancovici, B.:
Density correlations in the two-dimensional Coulomb gas.
J. Stat. Phys. {\bf 106}, 323--355 (2002)

\bibitem{Samaj04} \v{S}amaj, L.:
Is the two-dimensional one-component plasma exactly solvable?
J. Stat. Phys. {\bf 117}, 131--158 (2004)

\bibitem{Samaj11} \v{S}amaj, L., Trizac, E.:
Counter-ions at charged walls: Two-dimensional systems.
Eur. Phys. J. E {\bf 34}, 20 (2011)

\bibitem{Samaj14} \v{S}amaj, L., Trizac, E.:
Counter-ions between or at asymmetrically charged walls: 2D free-fermion point.
J. Stat. Phys. {\bf 156}, 932--947 (2014)

\bibitem{Samaj15} \v{S}amaj, L.:
Counter-ions near a charged wall: Exact results for disc and planar geometries.
J. Stat. Phys. {\bf 161}, 227--249 (2015)

\bibitem{Samaj16} \v{S}amaj, L.:
Amplitude function of asymptotic correlations along charged wall
in Coulomb fluids.
J. Stat. Phys. {\bf 164}, 304--320 (2016)

\bibitem{Sari76} Sari, R.R., Merlini, D., Calinon, R.:
On the ground state of the one-component classical plasma.
J. Phys. A: Gen. Phys. {\bf 9}, 1539--1551 (1976)

\bibitem{Tellez99} Tell\'ez, G., Forrester, P.J.:
Exact finite-size study of the 2D OCP at $\Gamma=4$ and $\Gamma=6$.
J. Stat. Phys. {\bf 97}, 489--521 (1999)

\bibitem{Tellez01} Tell\'ez, G.:
Two-dimensional Coulomb systems in a disk with ideal dielectric boundaries.
J. Stat. Phys. {\bf 104}, 945--970 (2001)

\bibitem{Tellez12} Tell\'ez, G., Forrester, P.J.:
Expanded Vandermonde powers and sum rules for the two-dimensional
one-component plasma.
J. Stat. Phys. {\bf 148}, 824--855 (2012)

\end{thebibliography}
\end{document}